\newif\ifarxiv
\arxivtrue

\ifarxiv
    \documentclass[paper=letter,parskip=half]{scrartcl}
    \usepackage[top=3cm, bottom=3cm, left=3.5cm, right=3.5cm]{geometry}
    \usepackage[affil-it]{authblk}
\else
    \documentclass[runningheads]{llncs}
\fi
\usepackage[utf8]{inputenc}
\usepackage[T1]{fontenc}
\usepackage{lmodern}
\usepackage{amssymb}
\usepackage{bm}
\usepackage{mathtools}
\usepackage{listings}
\usepackage{url}
\usepackage[protrusion=false]{microtype}

\def\N{\mathbb{N}}

\def\R{\mathbb{R}}
\def\Q{\mathbb{Q}}
\def\NN{\mathcal{N}}
\def\PP{\mathcal{P}}

\newcommand{\vect}[1]{\ensuremath{\bm #1}}

\newcommand{\AND}{\wedge}
\newcommand{\OR}{\vee}
\newcommand{\NOT}{\neg}
\newcommand{\ptcut}{\textsf{PtCut}}
\newcommand{\smtcut}{\textsf{SMTcut}}
\newcommand{\odebase}{\textsf{ODEbase}}
\newcommand{\odeparse}{\textsf{ODEparse}}
\def\eps{\varepsilon}

\newcommand\Tstrut{\rule{0pt}{2.6ex}}         

\ifarxiv
    \newcommand{\email}[1]{\texttt{#1}}
\fi

\begin{document}

\title{Computing Tropical Prevarieties with \\ Satisfiability Modulo Theories (SMT) Solvers}
\ifarxiv
    \author{Christoph Lüders\\ \email{chris@cfos.de}}
    \affil{Universität Bonn, Institut für Informatik II\\ Endenicher Allee 19a, 53115 Bonn, Germany}
    \date{June 26th, 2020}
\fi

\maketitle

\begin{abstract}
A novel way to use SMT (Satisfiability Modulo Theories) 
solvers to compute the tropical prevariety
(resp.\ equilibrium) of a polynomial system is presented.  The new method is benchmarked against
a naive approach that uses purely polyhedral methods.
It turns out that the SMT approach is faster than the polyhedral approach 
for models that would otherwise take more than one minute to compute, 
in many cases by a factor of 60 or more, and in the worst case is only slower by a factor of two.
Furthermore, the new approach is an anytime algorithm, thus offering a way to compute parts of the solution
when the polyhedral approach is infeasible.
\end{abstract}

\section{Introduction}

Tropical geometry \cite{brugall2013bit} has been used to find the order of time scales
of variables in chemical reaction networks \cite{SamalGrigorievFroehlichWeberRadulescu2015}
and for model reduction.
It has applications in economics and optimizations like network flows and scheduling.

\medskip
Satisfiability Modulo Theories (SMT) checking \cite{monniaux2016survey} 
is usually built on top of SAT (Boolean satisfiability),
which is the first problem that was proved, in the form of 3SAT, to be NP-complete.
SMT checking allows one to test a logical formula with unknowns and relations for
satisfiability and, if it is so, for an assignment of the unknowns that leads to
the formula's satisfiability.
SMT checking is used today in verification of computer hardware and software
and has advanced much in recent years due to advances in technology and 
industrial applications \cite{de2011satisfiability}.

\medskip
We present a novel approach to use SMT checking to compute the tropical
equilibrium (resp.\ prevariety).  We believe this to be of use,
since SMT is a very active field of research, yet 
problems of tropical geometry have not been solved with SMT until now.

In the following, we describe the idea of tropical geometry 
and SMT in the remainder of this section.
In Sect.~\ref{problem} we describe the exact problem
and Sect.~\ref{procedure} describes the proposed solution
and several possible improvements. Section~\ref{benchmarks}
has results of speed tests of our implementation \smtcut\ for
various SMT solvers and possible optimizations.

\subsection{Some Tropical Geometry}
\label{tropicalization}

Given a system of polynomial equations with zero r.h.s.,
the basic idea is to express its indeterminates $x_i \in \R_+$ and 
parameters $k_\alpha \in \R_+$ 
as powers of some $\eps \in (0,1)$ times a value $\bar{x}_i$ resp.\ $\bar{k}_\alpha$ that is roughly 1.
That is, indeterminates become $x_i = \bar{x}_i \eps^{a_i}$ and 
parameters $k_\alpha = \bar{k}_\alpha \eps^{\gamma_\alpha}$, 
with $\bar{x}_i \approx 1 \approx \bar{k}_\alpha$.

Let
$\vect{x} \in \R^d_+$ and $\vect{\alpha}$, $\vect{\beta} \in \N_0^m$ be multi-indices with
$\vect{x}^{\vect{\alpha}} = \prod_i x_i^{\alpha_i}$. 
Furthermore, let $\PP$ and $\NN$ be index sets of the terms 
with positive resp.\ negative sign.
We write the polynomial equation system as
\begin{equation}
    \label{classic:start}
    \sum_{\vect{\alpha \in \PP}} k_{\vect{\alpha}} \vect{x}^{\vect{\alpha}}
    - \sum_{\vect{\beta \in \NN}} k_{\vect{\beta}} \vect{x}^{\vect{\beta}} = 0. 
\end{equation}

Next, we sort all
terms with positive sign to one side and all terms with negative sign to the other.
Hence, the equation becomes 
\begin{equation}
    \label{classic:ss} 
    \sum_{\vect{\alpha \in \PP}} k_{\vect{\alpha}} \vect{x}^{\vect{\alpha}} =
    \sum_{\vect{\beta \in \NN}} k_{\vect{\beta}} \vect{x}^{\vect{\beta}}. 
\end{equation}

The critical observation is now that on each side almost all of the time one term
dominates all others \cite{Viro2000}. In tropical geometry we only look at this
dominating term.  A \emph{tropical root} is found when the dominating terms
on both sides cancel each other out.
Since $\eps < 1$, the inequality
$\bar{x} \eps^a > \bar{x}' \eps^{a'}$ is approximately equivalent to
$a < a'$. ``Domination'' thus means that $a$ is minimal. 

With the above sketched idea of \emph{tropicalization}, we transform
equation \eqref{classic:ss} into its tropical counterpart to look for tropical
roots: 
\[
    \min_{\vect{\alpha \in \PP}} \Big( \log_\eps (k_{\vect{\alpha}}) + \sum_i \alpha_i x_i \Big) =
    \min_{\vect{\beta \in \NN}} \Big( \log_\eps (k_{\vect{\beta}}) + \sum_i \beta_i x_i \Big).  
\]

In order for this equation to hold, the minimum has to be attained at least twice,
one time on each side.
Observe that the equation consists now only of minima of linear functions
and can thus readily be expressed as a set of polyhedra.

One polyhedron is defined by each combination of $\vect{\alpha} \in \PP$ and $\vect{\beta} \in \NN$
that yields a hyperplane via 
\begin{equation}
    \label{trop:eqn}       
    \log_\eps (k_{\vect{\alpha}}) + \sum_i \alpha_i x_i = \log_\eps (k_{\vect{\beta}}) + \sum_i \beta_i x_i,     
\end{equation}
while for all $\vect{\eta} \in \PP \cup \NN$ half-spaces are defined by
\begin{equation}
    \label{trop:ieq}
    \log_\eps (k_{\vect{\alpha}}) + \sum_i \alpha_i x_i \le \log_\eps (k_{\vect{\eta}}) + \sum_i \eta_i x_i. 
\end{equation}
A set of polyhedra is defined by cycling over all possible choices for 
$\vect{\alpha}$ and $\vect{\beta}$.

Since a polynomial system may consist of multiple equations, we get multiple sets
of polyhedra.  Because we are looking for solutions where all polynomials are zero
at the same time, all constraints \eqref{trop:eqn} and \eqref{trop:ieq} 
have to hold at the same time and hence the 
sets of polyhedra have to be intersected. The resulting set of polyhedra
is called a \emph{tropical equilibrium}.

For computations, the parameter $\eps$ in the above construction has 
to be provided. Furthermore, the $\log_\eps(k_{\alpha})$
are rounded to rationals. Tropical geometry focuses on the dominant terms
and hence yields only approximate results.  The resulting
coarse-graining can be helpful to find broad areas
where the dominant terms cancel each other out.

Equations like \eqref{classic:start} can be used to model chemical reaction
networks, where all quantities (concentrations and reaction speeds)
are positive values.  The resulting matching of positively
and negatively signed terms in \eqref{classic:ss} is 
called \emph{opposite sign condition}.

If no such distinction is made and all terms---regardless of their sign---%
can cancel each other out,
then a \emph{tropical prevariety} is computed. In the above formalism
this can be achieved by defining 
both $\PP$ and $\NN$ to contain the indices of all terms.
The tropical prevariety is a superset of the tropical equilibrium.

We will continue to speak of equilibria, but the presented algorithm works 
for prevarieties as well, since the difference is only in the input.
In Sect.~\ref{benchmarks} we show run-times only
for computations of equilibria, since the inputs for these benchmarks 
come from chemical reaction networks.

\subsection{Bringing SMT into the Picture}

The problem that we are solving is the intersection of several unions of polyhedra. 
That is, given polyhedra $P_{ij}$ and 
unions of (convex) polyhedra $B_i = P_{i1} \cup P_{i2} \cup \ldots \cup P_{in_i}$,
we are interested in the intersection $\bigcap_i B_i = \{ R_1, R_2, \ldots, R_\ell \}$,
where the $R_i$ are again (convex) polyhedra.
In this article we show how to use SMT checking to solve this problem.

Satisfiability Modulo Theories (SMT) checking allows us to decide if a logical
formula, with atoms that are themselves equations or inequalities,
is satisfiable or not.  
For example, $x > 1 \AND x < 2$ is such an SMT formula.
One has to specify a \emph{theory} of numbers that unknowns in the formula
can assume. In the above example, the problem is satisfiable in the theory of real numbers, 
but not in the theory of integers.
If an SMT problem is satisfiable, SMT can return a \emph{model},
which is an assignment for all unknowns in the formula.

SMT solvers may be used in \emph{incremental mode}, where one can
add additional \emph{assertions}, i.e., clauses that are combined with \texttt{AND}, 
and continue to look for further models after one has found a solution. 
This can save a lot of time and we will make use of it later.

\section{The Problem}
\label{problem}

A (convex) polyhedron is defined as the intersection
of finitely many hyperplanes and half-spaces.
Furthermore, each hyperplane can be expressed as two (closed) half-spaces, 
thus a polyhedron can
be described as a finite number of half-spaces \cite{Ziegler1995}.

Given the ambient space $\Q^d$, a (closed) half-space $H$ is a set
\begin{equation}
    \label{halfspace}
    H = \{ \vect{x} \in \Q^d \mid \lambda_0 + \lambda_1 x_1 + \lambda_2 x_2 + \ldots + 
    \lambda_d x_d \le 0, \lambda_i \in \Q \}. 
 \end{equation}
Given half-spaces $H_k$, we define a polyhedron as
\begin{equation}
    \label{polyhedron}
    P = \bigcap_k H_k.  
\end{equation}
A \emph{bag} is what we call a union of polyhedra $P_j$, i.e.,
\[
B = \bigcup_j P_j.
\]
Finally, we are looking for the intersection of said bags $B_i$, that is
\begin{equation}
    \label{result}
    V = \bigcap_i B_i = \bigcap_i \bigcup_j P_{ij} = \bigcup_k R_k.  
\end{equation}

The naive solution to the problem of computing the intersection is to cycle
successively through all combinations.  To do that, pick two 
bags $B_{j}$ and $B_{j'}$, $j \not= j'$, 
and intersect all polyhedra from one with all polyhedra from the other
to form a new bag $B'$.  Then, remove $B_j$ and $B_{j'}$ from the set 
of bags and insert $B'$ instead. Continue this procedure until 
there is only one bag left, which will then consist of the sought
polyhedra $R_1, R_2, \ldots, R_\ell$.
This is the solution that was used in \cite{SamalGrigorievFroehlichWeberRadulescu2015} 
and, with some refinements,
in \ptcut\ \cite{LuedersPtCut}.

The problem with this solution is that the complexity is exponential in the
number of bags.  In practice, it often happens that the number of intermediate
results, i.e., the number of polyhedra in some $B'$, is very high, even if in the 
end there are only a few solution polyhedra.  This \emph{intermediate expression swell}
makes computing the intersection $V$ infeasible for some models.

Table \ref{tab:comp} contains BioModels from our survey and lists their number of
resulting polyhedra and maximum number of intermediate polyhedra.
Details on the computation can be found in Sect.~\ref{benchmarks}.

\section{The Procedure}
\label{procedure}

First, we have to formulate our problem as an SMT problem.
Fortunately, it is easy to convert a polyhedron as defined 
in \eqref{polyhedron} into a logical formula.
Set theory maps easily to logical formulas with union
mapping to logical \texttt{OR} and intersection to logical \texttt{AND}.
In the following, $\tilde{H}$ denotes the logical formula
that defines the set $H$.
Thus, \eqref{result} expands to 
\[
    \tilde{V} = \bigwedge_i \bigvee_j \bigwedge_k \tilde{H}_{ijk}  
\]
and definition \eqref{halfspace} of $H_{ijk}$ employs a linear function that can be used 
as a formula in SMT. Call the resulting SMT formula $f$.
We can use an SMT solver to decide the satisfiability of $f$
and, what's more important, get an $\vect{x} \in \Q^d$ that satisfies 
the constraints, if there is one.

Next, we look for a matching polyhedron that includes $\vect{x}$ and is included 
in the solution $V$. Since $\vect{x}$ is contained in the intersection of 
the $B_i$, it must be contained in at least one polyhedron $P_{ij}$ per bag $B_i$.
Thus, we cycle through all $B_i$ to find a containing $P_{ij}$,
call it $P'_i$.  (There may be more than one $P_{ij} \ni \vect{x}$,
but any will do.)

Obviously, the intersection $R' = \bigcap_i P'_i$
includes $\vect{x}$,
but most likely $R'$ has higher dimension than that.  Furthermore, since $R'$ is the 
intersection of exactly one polyhedron per bag it is included in $V$ as well.
Hence, we have found a polyhedron that includes $\vect{x}$.

In the next step, we modify our initial formula $f$ to exclude the polyhedron $R'$,
like this:
\[
    f' = f \AND \NOT R'.  
\]
Notice that we are only adding another assertion to the formula, so we can 
utilize the incremental mode of SMT solvers to save (a lot of!) time for its next computation.

The important observation here is that we are expanding the original formula
$f$---which describes all solution points---to exclude what we already know
to be a solution and continue the search.  
Thus the procedure generates an ever growing subset of the solution $V$,
making it an \emph{anytime algorithm} \cite{zilberstein1996using}.
We can iterate this process until formula $f'$ is unsatisfiable.

\medskip
\noindent This is the algorithm in Python-style pseudocode:
\lstset{language=Python, basicstyle=\scriptsize\ttfamily, frame=L, numbers=left}
\begin{lstlisting}
# input: a list 'll' of sets of polyhedra.
# output: a list 'rr' of polyhedra.
def compute_polyhedron_dnf(ll):
    # set the solver to re-use its heuristics
    solver = Solver(incremental=True)
    f = convert_to_SMT_formula(ll)
    rr = []             # results list
    while True:
        # add the formula to the (existing) assertions.
        solver.add_assertion(f)
        # get the model (a variable assignment) that fits the 
        # constraints, or None if 'f' is unsatisfiable.
        x = solver.get_model()
        if x is None:   
            break       # formula unsatisfiable
        R = []          # resulting polyhedron
        # cycle through all bags 'B' and 
        # collect constraints of polyhedron containing 'x'.
        for B in ll:
            # cycle through all polyhedra 'P' in bag 'B'.
            for P in B:
                if P.contains(x):
                    R.append(P.constraints())
                    break
        # now 'R' defines a polyhedron surely in the intersection.
        # exclude 'R' from further searches.
        f = Not(R)      # new assertion for next round
        rr.append(R)
    return rr           # list of polyhedra.
\end{lstlisting}

The result of this function is a list of polyhedra. 
Mathematically, this union of polyhedra describes the equilibrium (resp.\ prevariety) $V$.
Yet, there are some problems that we address in the next section.

The logic used for SMT formulas is \texttt{QF\_LRA}, that is,
quantifier-free linear real arithmetic (here ``real'' means rational). 
This allows Boolean propositional logic
of equations/inequalities consisting of linear functions over
elements of $\Q$ \cite{barrett2010smt}.

\subsection{Improvements to the Procedure}

\subsubsection*{Non-maximal Polyhedra}

The main issue we experience with the procedure \texttt{compute\_polyhedron\_dnf}
is that the polyhedron $R$ computed from point $\vect{x}$
is often not maximal.  That is, $R$ is only a lower dimensional face of a
higher-dimensional polyhedron.  The full high-dimensional polyhedron
will eventually be found by the procedure, but earlier-found lower-dimensional
faces would still remain in the result list \texttt{rr}, albeit superfluous.

To avoid this, we test if each newly found polyhedron $R$ 
is included in one of the already computed polyhedra of result list \texttt{rr}.
Unfortunately, this causes quadratic run-time in the number of resulting
polyhedra. 
But there is an observation that can reduce the constant.

If the newly found polyhedron $R$ is included in some already found polyhedron $R'$,
then obviously, point $\vect{x} \in R$ is included in $R'$ as well.
Testing if a point is included in a polyhedron is simple and fast, so
one can test this first. Only if this test succeeds one must perform
the full polyhedron inclusion test.  Measurements show that with this
heuristic, almost all polyhedron inclusion tests can be avoided.
See Sect.~\ref{benchmarks} for details.

In our procedure, we would have to modify function \texttt{contains} in
line 22 and function \texttt{append} in line 23 according to these observations.

\subsubsection*{Superfluous Constraints}

Another issue is the redundancy of the constraints that are collected in
line~23.  Efficiency can be increased by minimizing the set of constraints:
the larger the number of constraints, the larger the memory demand and,
of course, SMT checking times.  

One can simply cycle through all constraints, test if each of them is 
really required and if not, drop it.  
The remaining set is not necessarily a minimal set, though.

Here's how this can be done:
Let $c$ be the constraint in consideration, 
$g$ the formula before and $g'=g \AND c$ after the addition.
If $g'$ is more restrictive than $g$ (i.e., $c$ makes a difference),
then the following is unsatisfiable:
\begin{align*}
g \AND \NOT g' 
   &= g \AND \NOT (g \AND c)  \\
   &= g \AND (\NOT g \OR \NOT c)  \\
   &= (g \AND \NOT g) \OR (g \AND \NOT c)  \\
   &= g \AND \NOT c.
\end{align*}
We use SMT checking and apply this formula to all constraints to 
drop superfluous ones.

\subsubsection*{Preprocessing}

We explored the possibility to improve the speed of the procedure by
preprocessing the input, i.e., the sets of polyhedra.  

For one, one can collect all constraints from all bags with only
one polyhedron each.  Call the resulting polyhedron $C$.
Because of distributivity these constraints
hold for all polyhedra of the solution.  Hence, we can intersect
all polyhedra in their bags with $C$ to test if the intersection is empty,
in which case we drop the polyhedron 
from its bag to reduce computation time.

A more powerful version of this technique can be used to test 
polyhedra in all bags on if they are required for 
the definition of the solution in \eqref{result}.
Let $B$ be the polyhedron in question, $A$ the union of all 
other polyhedra in $B$'s bag and $C$ the intersection of all other bags.
Then the solution is $(A \cup B) \cap C$.  If $B$ is required,
then $A \cap C \not= (A \cup B) \cap C$ and
in particular $A \cap C \subsetneq (A \cup B) \cap C$.
Thus, the following set is non-empty: 
\begin{align*}
((A \cup B) \cap C) \setminus (A \cap C) 
  &= (A \cup B) \cap C \cap \overline{A \cap C} \\
  &= (A \cup B) \cap C \cap (\overline{A} \cup \overline{C}) \\
  &= (A \cup B) \cap C \cap \overline{A} \\
  &= B \cap C \cap \overline{A}.
\end{align*}
This can easily be tested with SMT checking
for each polyhedron $B$ per bag and the superfluous polyhedra are dropped,
again reducing the computation time.

\section{Benchmarks}
\label{benchmarks}

To benchmark the procedure \texttt{compute\_polyhedron\_dnf}, we created
input sets of polyhedra of chemical reaction networks.  The reaction networks
were taken from the BioModels database \cite{BioModels2006}, which contains
models formulated in SBML \cite{Hucka2003}.
The SBML models were converted to ODE systems with polynomial or rational functions using
\odeparse\ \cite{LuedersODEparse}.  
The resulting 
ODE systems can be downloaded from \odebase\ \cite{LuedersErramiNeidhardtSamalWeber2019}
at \url{http://odebase.cs.uni-bonn.de}.
If the ODE systems contained rational functions,
we multiplied each equation with its common denominator
to get a polynomial.

These polynomials were then tropicalized as sketched in 
Sect.~\ref{tropicalization}.
The parameter $\eps$ was set to $1/11$ and the logarithms 
in \eqref{trop:eqn} and \eqref{trop:ieq} were rounded to integers.
The sets of polyhedra created by tropicalization were saved
with polyhedra expressed as sets of equalities and inequalities.

The software used for polyhedral computation was \ptcut~3.5.1 \cite{LuedersPtCut}
and was run under Windows~10 64-bit using Python~3.7.7.
Polyhedral computation were done with the help of the
Parma Polyhedra Library (PPL) \cite{BagnaraHillZaffanella2008},
version~1.2.

The procedure \texttt{compute\_polyhedron\_dnf} was implemented in Python as
\smtcut\ and \smtcut\ 4.6.4 was used for benchmarking.
\smtcut\ is available under a free software license
from \url{https://gitlab.com/cxxl/smtcut}.  

Neither \ptcut\ nor \smtcut\ make active use of multithreading.
All input and output files are available
from \url{https://gitlab.com/cxxl/smtcut_data_2}
in one large ($\approx$~42~MiB) repository.

The machine for tests was an Intel Core i7-5820K CPU with 48~GB of memory.
We used the solver-agnostic framework PySMT~0.9.0 \cite{pysmt2015}
to access SMT solvers.  Of the seven supported solvers, 
the solvers CUDD, PicoSAT and Boolector could not be used, since
they do not support the \texttt{QF\_LRA} logic.
This left us with
the SMT solvers MathSAT~5.6.1 \cite{mathsat5}, Z3~4.8.7 \cite{z3smt}, 
CVC4~1.7-prerelease \cite{CVC4} and Yices~2.6.2 \cite{yicessmt}.
\smtcut\ was run on Ubuntu Linux 18.04 64-bit with Python~3.7.6.


Table~\ref{tab:solvers} shows a comparison of run-times for all
BioModels that completed within two hours, sorted by ascending
run-time.

\begin{table}[!h]
    \caption{Comparison of run-times for different solvers sorted by ascending run-time.
    The minimal time is printed in bold. Column ``BM'' contains the BioModel number;
    ``Rounds'' is the number of iterations needed to complete the procedure.}
    \label{tab:solvers}
    
    \begin{center}
        {\scriptsize
        \begin{tabular}{r|rrrr|rrrr|rrrr}
            & \multicolumn{4}{c}{Run-time [s]} & \multicolumn{4}{|c}{Rounds} & \multicolumn{4}{|c}{Run-time / round [s]}  \\
            BM & MSAT & Z3 & CVC4 & Yices & MSAT & Z3 & CVC4 & Yices & MSAT & Z3 & CVC4 & Yices  \\
            \hline\Tstrut
            & \hphantom{\textbf{0000.0}} & \hphantom{\textbf{0000.0}} & \hphantom{\textbf{0000.0}} & \hphantom{\textbf{0000.0}} & \hphantom{\textbf{00000}} & \hphantom{\textbf{00000}} & \hphantom{\textbf{00000}} & \hphantom{\textbf{00000}} & \hphantom{\textbf{0000.0}} & \hphantom{\textbf{0000.0}} & \hphantom{\textbf{0000.0}} & \hphantom{\textbf{0000.0}} \\[-1em]
            397 &      0.1 &      0.1 &      0.1 &  \textbf{    0.1} &  \textbf{     1} &  \textbf{     1} &  \textbf{     1} &  \textbf{     1} &      0.128 &      0.108 &      0.147 &  \textbf{    0.092}  \\
 26 &      0.1 &      0.1 &      0.1 &  \textbf{    0.1} &  \textbf{     7} &       8 &       8 &       8 &      0.016 &      0.017 &      0.016 &  \textbf{    0.012}  \\
 41 &      0.2 &      0.2 &      0.2 &  \textbf{    0.1} &  \textbf{     7} &      10 &  \textbf{     7} &       8 &      0.025 &      0.022 &      0.030 &  \textbf{    0.017}  \\
 30 &      0.2 &      0.2 &      0.2 &  \textbf{    0.2} &       8 &       8 &  \textbf{     7} &       8 &      0.026 &      0.028 &      0.029 &  \textbf{    0.020}  \\
492 &      0.3 &      0.2 &      0.3 &  \textbf{    0.2} &  \textbf{     2} &  \textbf{     2} &  \textbf{     2} &  \textbf{     2} &      0.141 &      0.120 &      0.158 &  \textbf{    0.083}  \\
491 &      0.3 &      0.2 &      0.3 &  \textbf{    0.2} &  \textbf{     2} &  \textbf{     2} &  \textbf{     2} &  \textbf{     2} &      0.146 &      0.118 &      0.130 &  \textbf{    0.084}  \\
 48 &      0.8 &      0.7 &      0.8 &  \textbf{    0.4} &      12 &      13 &  \textbf{     8} &  \textbf{     8} &      0.068 &      0.054 &      0.096 &  \textbf{    0.044}  \\
482 &      0.5 &      0.8 &      0.6 &  \textbf{    0.4} &  \textbf{    19} &      25 &      21 &      20 &      0.028 &      0.032 &      0.027 &  \textbf{    0.020}  \\
 28 &      0.6 &      0.9 &  \textbf{    0.4} &      0.5 &      25 &      35 &  \textbf{    20} &      32 &      0.022 &      0.025 &      0.022 &  \textbf{    0.017}  \\
637 &      0.6 &      0.7 &      0.6 &  \textbf{    0.5} &      21 &      27 &  \textbf{    19} &      24 &      0.026 &      0.027 &      0.031 &  \textbf{    0.020}  \\
221 &      0.7 &      0.8 &      0.7 &  \textbf{    0.6} &      55 &      58 &  \textbf{    53} &      68 &      0.012 &      0.014 &      0.013 &  \textbf{    0.009}  \\
 21 &      0.7 &      0.9 &      0.7 &  \textbf{    0.6} &  \textbf{    56} &      65 &      61 &      67 &      0.012 &      0.015 &      0.011 &  \textbf{    0.009}  \\
315 &      0.8 &      1.0 &  \textbf{    0.8} &      0.8 &      25 &      30 &  \textbf{    19} &      36 &      0.031 &      0.033 &      0.041 &  \textbf{    0.022}  \\
328 &      2.0 &      2.9 &      1.7 &  \textbf{    1.5} &      93 &     105 &  \textbf{    89} &      90 &      0.022 &      0.027 &      0.019 &  \textbf{    0.017}  \\
200 &      2.0 &      1.9 &      2.2 &  \textbf{    1.6} &      38 &      41 &  \textbf{    34} &      45 &      0.053 &      0.048 &      0.063 &  \textbf{    0.035}  \\
599 &      2.1 &      2.3 &  \textbf{    1.6} &      1.7 &      48 &      48 &  \textbf{    36} &      52 &      0.043 &      0.048 &      0.045 &  \textbf{    0.033}  \\
 22 &      2.2 &      3.5 &  \textbf{    1.8} &      2.3 &     199 &     235 &  \textbf{   165} &     259 &      0.011 &      0.015 &      0.011 &  \textbf{    0.009}  \\
666 &      2.7 &      3.0 &  \textbf{    2.2} &      2.6 &      93 &  \textbf{    85} &      85 &     111 &      0.029 &      0.038 &      0.026 &  \textbf{    0.023}  \\
222 &      3.2 &      4.0 &      3.1 &  \textbf{    2.3} &     246 &     257 &  \textbf{   228} &     230 &      0.013 &      0.016 &      0.014 &  \textbf{    0.010}  \\
638 &      2.6 &      2.9 &      3.2 &  \textbf{    2.4} &      37 &      46 &  \textbf{    29} &      51 &      0.071 &      0.062 &      0.109 &  \textbf{    0.047}  \\
489 &      4.3 &      4.9 &      3.0 &  \textbf{    2.5} &      72 &      82 &  \textbf{    49} &      54 &      0.059 &      0.060 &      0.061 &  \textbf{    0.046}  \\
365 &      4.6 &      5.1 &      4.3 &  \textbf{    3.6} &      80 &      82 &  \textbf{    74} &      82 &      0.058 &      0.062 &      0.059 &  \textbf{    0.044}  \\
396 &  \textbf{    3.9} &      4.8 &      4.2 &      5.0 &  \textbf{    75} &      82 &      86 &     121 &      0.053 &      0.059 &      0.049 &  \textbf{    0.041}  \\
147 &      4.1 &      6.0 &      4.4 &  \textbf{    4.0} &  \textbf{    84} &     125 &      90 &     117 &      0.049 &      0.048 &      0.049 &  \textbf{    0.035}  \\
230 &      5.5 &      7.0 &  \textbf{    5.1} &      5.3 &     115 &     143 &  \textbf{    99} &     139 &      0.048 &      0.049 &      0.052 &  \textbf{    0.038}  \\
498 &      6.5 &      9.1 &  \textbf{    6.1} &      6.9 &  \textbf{   231} &     285 &     242 &     333 &      0.028 &      0.032 &      0.025 &  \textbf{    0.021}  \\
431 &      7.7 &      9.7 &  \textbf{    6.6} &      7.4 &     199 &     224 &  \textbf{   183} &     243 &      0.039 &      0.043 &      0.036 &  \textbf{    0.030}  \\
105 &      9.6 &      9.4 &      9.7 &  \textbf{    6.7} &  \textbf{   141} &     153 &     143 &     152 &      0.068 &      0.061 &      0.068 &  \textbf{    0.044}  \\
102 &     14.8 &     19.1 &     12.2 &  \textbf{   11.8} &     399 &     538 &  \textbf{   395} &     497 &      0.037 &      0.036 &      0.031 &  \textbf{    0.024}  \\
 32 &     22.0 &     23.1 &  \textbf{   12.8} &     23.5 &     418 &     388 &  \textbf{   282} &     504 &      0.053 &      0.060 &  \textbf{    0.046} &      0.047  \\
407 &     20.7 &     30.2 &  \textbf{   15.7} &     55.0 &     307 &     389 &  \textbf{   261} &     881 &      0.067 &      0.078 &  \textbf{    0.060} &      0.062  \\
477 &     42.0 &     72.0 &  \textbf{   33.8} &     49.3 &     793 &     940 &  \textbf{   745} &    1051 &      0.053 &      0.077 &  \textbf{    0.045} &      0.047  \\
576 &     55.1 &     78.0 &  \textbf{   48.6} &     62.9 &  \textbf{   838} &    1064 &     881 &    1076 &      0.066 &      0.073 &  \textbf{    0.055} &      0.058  \\
 93 &    124.4 &    121.0 &  \textbf{   55.3} &    139.7 &    1390 &    1345 &  \textbf{   870} &    1865 &      0.090 &      0.090 &  \textbf{    0.064} &      0.075  \\
183 &  \textbf{   69.8} &    132.0 &     77.7 &    115.1 &      29 &     194 &  \textbf{    14} &     284 &      2.408 &      0.681 &      5.551 &  \textbf{    0.405}  \\
501 &    116.1 &    192.2 &  \textbf{  102.8} &    352.5 &  \textbf{  1018} &    1516 &    1045 &    2819 &      0.114 &      0.127 &  \textbf{    0.098} &      0.125  \\
430 &    209.8 &    196.2 &  \textbf{  128.2} &    405.5 &    2321 &    2448 &  \textbf{  2142} &    4798 &      0.090 &      0.080 &  \textbf{    0.060} &      0.085  \\
103 &    231.0 &    449.2 &  \textbf{  224.9} &    337.2 &    2781 &    3072 &  \textbf{  2609} &    2923 &  \textbf{    0.083} &      0.146 &      0.086 &      0.115  \\
 74 &   1165.6 &   1118.5 &  \textbf{  941.5} &   1694.2 &   14227 &  \textbf{ 13352} &   13352 &   23245 &      0.082 &      0.096 &  \textbf{    0.071} &      0.073  \\
 73 &   1087.1 &   1647.9 &  \textbf{ 1067.7} &   1485.4 &  \textbf{ 14977} &   16294 &   15522 &   21118 &      0.073 &      0.101 &  \textbf{    0.069} &      0.070  \\
 61 &   3045.7 &   4084.1 &  \textbf{ 1661.3} &   4187.1 &  \textbf{ 12564} &   16326 &   12915 &   24078 &      0.242 &      0.250 &  \textbf{    0.129} &      0.174  \\    
        \end{tabular}
        }
    \end{center}
\end{table}

The lowest computation times were achieved with CVC4, if%
---with the exception of BioModel~183---computation took more than 12~seconds.
Below that time Yices was often faster.
If we look at run-time per round we see that Yices had always
the lowest run-times per round for models that took less than 12~seconds
to compute.  For unknown reasons, Yices often needs more
rounds than other solvers, especially CVC4. 

The geometric means of run-times of completed computations 
for CVC4, Yices, MathSAT and Z3 were
4.93, 5.37, 5.49 and 6.77 seconds, respectively.
For that reason, we used CVC4 as solver in further comparisons.

\medskip
Table~\ref{tab:comp} shows a comparison of run-times between
\ptcut\ and \smtcut\ with CVC4 as solver.
We can make several observations from the data.

\begin{table}[!ht]        
    \caption{Comparison of run-times for \smtcut\ and \ptcut\ sorted by ascending \smtcut\ run-time.
    Minimal run-times are set in bold.
    Columns ``BM'', ``Dim'', ``Comb'', ``PH'', and ``IntMaxPH'' contain 
    the BioModel number, dimension, theoretic number of combinations, 
    number of polyhedra in the solution, and maximal intermediate polyhedra, 
    respectively. A star in column ``R'' signifies a 
    model with a rational vector field.}
    \label{tab:comp}
    
    \begin{center}
        {\scriptsize
        \begin{tabular}{rrrrrr|rrr}            
            BM & R & Dim & Comb & PH & IntMaxPH & \smtcut & \ptcut & Speed-up \\
            \hline\Tstrut
             26 &    &  10 &  $10^{14}$ &  6     &       32 &      0.1 &  \textbf{    0.1}  &  1.0  \\
397 &  * &  10 &  $10^{ 8}$ &  0     &  --- &  \textbf{    0.1} &     14.9  &  101.6  \\
 30 &    &  16 &  $10^{22}$ &  6     &       48 &  \textbf{    0.2} &      0.3  &  1.7  \\
 41 &    &  11 &  $10^{10}$ &  4     &      924 &  \textbf{    0.2} &      1.4  &  6.8  \\
491 &    &  36 &  $10^{ 6}$ &  1     &        1 &  \textbf{    0.3} &     17.2  &  66.6  \\
492 &    &  18 &  $10^{ 9}$ &  1     &        1 &  \textbf{    0.3} &      2.9  &  9.1  \\
 28 &    &  22 &  $10^{10}$ &  17    &      119 &      0.4 &  \textbf{    0.4}  &  1.0  \\
482 &  * &  23 &  $10^{26}$ &  17    &      495 &  \textbf{    0.6} &      2.3  &  4.1  \\
637 &    &  16 &  $10^{25}$ &  12    &     1140 &  \textbf{    0.6} &      4.2  &  7.1  \\
221 &  * &  19 &  $10^{20}$ &  50    &     2573 &  \textbf{    0.7} &      2.1  &  3.1  \\
 21 &  * &  33 &  $10^{84}$ &  46    &     3408 &  \textbf{    0.7} &      3.0  &  4.3  \\
 48 &  * &  10 &  $10^{21}$ &  5     &     1160 &  \textbf{    0.8} &     29.2  &  38.1  \\
315 &    &  13 &  $10^{14}$ &  13    &      432 &  \textbf{    0.8} &      2.5  &  3.3  \\
599 &    &  24 &  $10^{23}$ &  24    &      456 &  \textbf{    1.6} &      5.8  &  3.6  \\
328 &    &  17 &  $10^{23}$ &  86    &      140 &      1.7 &  \textbf{    1.4}  &  0.8  \\
 22 &  * &  67 &  $10^{38}$ &  147   &     1170 &  \textbf{    1.8} &      2.0  &  1.1  \\
200 &    &  27 &  $10^{18}$ &  20    &     4704 &  \textbf{    2.2} &     64.0  &  29.7  \\
666 &  * &  22 &  $10^{16}$ &  64    &      464 &  \textbf{    2.2} &      5.6  &  2.6  \\
489 &    &  24 &  $10^{16}$ &  42    &     4824 &  \textbf{    3.0} &     57.5  &  19.4  \\
222 &  * &  8 &  $10^{13}$ &  192   &    12516 &  \textbf{    3.1} &     11.0  &  3.5  \\
638 &    &  8 &  $10^{15}$ &  13    &     2124 &  \textbf{    3.2} &     32.3  &  10.2  \\
396 &  * &  18 &  $10^{ 9}$ &  54    &      972 &  \textbf{    4.2} &     21.7  &  5.2  \\
365 &    &  19 &  $10^{ 9}$ &  70    &    15030 &  \textbf{    4.3} &    583.7  &  134.2  \\
147 &    &  30 &  $10^{ 9}$ &  54    &     5069 &  \textbf{    4.4} &     18.7  &  4.2  \\
230 &    &  36 &  $10^{16}$ &  68    &     3330 &  \textbf{    5.1} &     24.6  &  4.8  \\
498 &  * &  50 &  $10^{17}$ &  214   &     1750 &      6.1 &  \textbf{    3.3}  &  0.5  \\
431 &    &  47 &  $10^{16}$ &  155   &      984 &  \textbf{    6.6} &     13.2  &  2.0  \\
105 &    &  27 &  $10^{28}$ &  130   &    21088 &  \textbf{    9.7} &    644.2  &  66.5  \\
102 &    &  27 &  $10^{21}$ &  322   &     4784 &     12.2 &  \textbf{    7.2}  &  0.6  \\
 32 &    &  43 &  $10^{11}$ &  244   &     1092 &  \textbf{   12.8} &    110.0  &  8.6  \\
407 &    &  23 &  $10^{16}$ &  212   &    15010 &  \textbf{   15.7} &    968.2  &  61.7  \\
477 &    &  35 &  $10^{24}$ &  467   &    23460 &  \textbf{   33.8} &    571.0  &  16.9  \\
576 &    &  57 &  $10^{10}$ &  756   &    10752 &  \textbf{   48.6} &     55.8  &  1.1  \\
 93 &    &  52 &  $10^{25}$ &  596   &    47772 &  \textbf{   55.3} &   6113.4  &  110.5  \\
183 &    &  19 &  $10^{ 9}$ &  1     &  --- &  \textbf{   77.7} &  >\,72000.0  &  >\,926.4  \\
501 &    &  35 &  $10^{25}$ &  916   &  --- &  \textbf{  102.8} &  >\,72000.0  &  >\,700.3  \\
430 &    &  34 &  $10^{18}$ &  1676  &     4683 &    128.2 &  \textbf{   58.0}  &  0.5  \\
103 &    &  30 &  $10^{16}$ &  1938  &   111402 &  \textbf{  224.9} &    763.5  &  3.4  \\
 74 &    &  12 &  $10^{12}$ &  9685  &  --- &  \textbf{  941.5} &  >\,72000.0  &  >\,76.5  \\
 73 &    &  21 &  $10^{24}$ &  13449 &  --- &  \textbf{ 1067.7} &  >\,72000.0  &  >\,67.4  \\
 61 &    &  34 &  $10^{27}$ &  10084 &  1784868 &  \textbf{ 1661.3} &  26408.6  &  15.9  \\
 14 &    &  86 &  $10^{67}$ &  >\,2663 &  --- &  \textbf{>\,7200.0} &  >\,72000.0  &  ---  \\
151 &    &  66 &  $10^{44}$ &  >\,7734 &  --- &  \textbf{>\,7200.0} &  >\,72000.0  &  ---  \\
410 &    &  53 &  $10^{41}$ &  >\,10612 &  --- &  \textbf{>\,7200.0} &  >\,72000.0  &  ---  \\
560 &    &  59 &  $10^{29}$ &  >\,12577 &  --- &  \textbf{>\,7200.0} &  >\,72000.0  &  ---  \\
730 &  * &  45 &  $10^{47}$ &  >\,15310 &  --- &  \textbf{>\,7200.0} &  >\,72000.0  &  ---  \\    
        \end{tabular}
        }
    \end{center}
\end{table}

\begin{itemize}
\item BioModels 14, 151, 410, 560 and 730:
each of them could not be computed with \ptcut\ and 
it is likely infeasible to do so. 
\smtcut\ was at least able to compute part of the solution,
even though it is unknown how large a part of the full solution this constitutes.

\item BioModels 183 and 491, 492: Here \smtcut\ was able to play out its
full potential: with only one polyhedron in the solution,
it took only some rounds until the maximal polyhedron was found.
On the other hand, the computation of BioModel~183 with \ptcut\
was terminated after 20~hours of work
with an intermediate number of 15000 polyhedra and still only
5 of 65 iterations done.

\item BioModel 397 does not have a solution
at all.  \smtcut\ find this in 0.1~s, whereas \ptcut\
takes almost 15~s.

\item BioModels 73, 74, 93, 105, 183, 365, 397, 407, 491, and 501:
computation time of \smtcut\ was at least 60 times lower than with
\ptcut.

\item BioModels 102, 328, 430 and 498: here \ptcut\ was up to 2 times
faster than \smtcut. The reason is that for models of dimension less than~20 
and with many polyhedra, \ptcut\ can be faster if the 
intermediate expression swell is not too large.

\item In this overview, \smtcut\ was the better choice in all cases where
\ptcut\ needed more than 58~s of computation time, and \smtcut\ never took more
than twice the time as \ptcut.
\end{itemize}

\subsection{Benchmarks of Preprocessing}

Table~\ref{tab:preproc} shows a comparison of times with and without
preprocessing the input.  
We make the following observations:

\begin{table}[!ht]
    \caption{Run-times in seconds without and with preprocessing for different 
    models using the CVC4 solver, sorted ascending by number of polyhedra. 
    Minimal times are in boldface.
    Column ``PP time'' contains the time for preprocessing and 
    column ``after PP'' the run-time minus the preprocessing time.}
    \label{tab:preproc}

    \begin{center}
        {\scriptsize
        \begin{tabular}{rrrr|rrrrr}
            BM & Dim & Comb & PH & No PP & With PP & PP time & After PP & Speed-up \\
            \hline\Tstrut
            397 &  10 &  $10^{ 8}$ &  0     &  \textbf{0.1} &  0.5 &  0.5 &  0.0 &  0.29  \\
183 &  19 &  $10^{ 9}$ &  1     &  \textbf{77.7} &  2988.7 &  2988.7 &  0.0 &  0.03  \\
491 &  36 &  $10^{ 6}$ &  1     &  \textbf{0.3} &  3.4 &  3.4 &  0.0 &  0.08  \\
492 &  18 &  $10^{ 9}$ &  1     &  \textbf{0.3} &  3.4 &  3.4 &  0.0 &  0.09  \\
 41 &  11 &  $10^{10}$ &  4     &  \textbf{0.2} &  2.6 &  2.6 &  0.1 &  0.08  \\
 48 &  10 &  $10^{21}$ &  5     &  \textbf{0.8} &  29.8 &  29.6 &  0.2 &  0.03  \\
 26 &  10 &  $10^{14}$ &  6     &  \textbf{0.1} &  0.9 &  0.8 &  0.1 &  0.14  \\
 30 &  16 &  $10^{22}$ &  6     &  \textbf{0.2} &  2.7 &  2.6 &  0.1 &  0.07  \\
637 &  16 &  $10^{25}$ &  12    &  \textbf{0.6} &  5.3 &  5.0 &  0.3 &  0.11  \\
315 &  13 &  $10^{14}$ &  13    &  \textbf{0.8} &  9.3 &  8.9 &  0.4 &  0.08  \\
638 &  8 &  $10^{15}$ &  13    &  \textbf{3.2} &  54.9 &  54.2 &  0.8 &  0.06  \\
 28 &  22 &  $10^{10}$ &  17    &  \textbf{0.4} &  2.5 &  2.0 &  0.4 &  0.18  \\
482 &  23 &  $10^{26}$ &  17    &  \textbf{0.6} &  2.8 &  2.5 &  0.4 &  0.20  \\
200 &  27 &  $10^{18}$ &  20    &  \textbf{2.2} &  29.2 &  28.3 &  0.9 &  0.07  \\
599 &  24 &  $10^{23}$ &  24    &  \textbf{1.6} &  9.7 &  8.4 &  1.2 &  0.17  \\
489 &  24 &  $10^{16}$ &  42    &  \textbf{3.0} &  59.9 &  57.6 &  2.3 &  0.05  \\
 21 &  33 &  $10^{84}$ &  46    &  \textbf{0.7} &  2.2 &  1.6 &  0.6 &  0.31  \\
221 &  19 &  $10^{20}$ &  50    &  \textbf{0.7} &  2.9 &  2.3 &  0.6 &  0.24  \\
147 &  30 &  $10^{ 9}$ &  54    &  \textbf{4.4} &  20.8 &  18.1 &  2.7 &  0.21  \\
396 &  18 &  $10^{ 9}$ &  54    &  \textbf{4.2} &  20.6 &  18.3 &  2.3 &  0.20  \\
666 &  22 &  $10^{16}$ &  64    &  \textbf{2.2} &  4.8 &  2.7 &  2.1 &  0.45  \\
230 &  36 &  $10^{16}$ &  68    &  \textbf{5.1} &  49.6 &  44.4 &  5.2 &  0.10  \\
365 &  19 &  $10^{ 9}$ &  70    &  \textbf{4.3} &  13.6 &  10.8 &  2.8 &  0.32  \\
328 &  17 &  $10^{23}$ &  86    &  \textbf{1.7} &  3.5 &  1.8 &  1.7 &  0.49  \\
105 &  27 &  $10^{28}$ &  130   &  \textbf{9.7} &  93.1 &  85.6 &  7.5 &  0.10  \\
 22 &  67 &  $10^{38}$ &  147   &  \textbf{1.8} &  3.7 &  1.8 &  2.0 &  0.48  \\
431 &  47 &  $10^{16}$ &  155   &  \textbf{6.6} &  18.2 &  11.8 &  6.4 &  0.36  \\
222 &  8 &  $10^{13}$ &  192   &  \textbf{3.1} &  6.7 &  3.7 &  3.0 &  0.47  \\
407 &  23 &  $10^{16}$ &  212   &  \textbf{15.7} &  27.4 &  10.9 &  16.5 &  0.57  \\
498 &  50 &  $10^{17}$ &  214   &  \textbf{6.1} &  8.6 &  3.2 &  5.4 &  0.71  \\
 32 &  43 &  $10^{11}$ &  244   &  \textbf{12.8} &  17.4 &  5.8 &  11.6 &  0.74  \\
102 &  27 &  $10^{21}$ &  322   &  \textbf{12.2} &  25.7 &  15.7 &  10.0 &  0.48  \\
477 &  35 &  $10^{24}$ &  467   &  33.8 &  \textbf{26.5} &  5.6 &  20.8 &  1.28  \\
 93 &  52 &  $10^{25}$ &  596   &  \textbf{55.3} &  86.2 &  39.0 &  47.2 &  0.64  \\
576 &  57 &  $10^{10}$ &  756   &  \textbf{48.6} &  62.0 &  20.0 &  42.0 &  0.78  \\
501 &  35 &  $10^{25}$ &  916   &  102.8 &  \textbf{87.0} &  13.2 &  73.9 &  1.18  \\
430 &  34 &  $10^{18}$ &  1676  &  128.2 &  \textbf{119.1} &  7.7 &  111.4 &  1.08  \\
103 &  30 &  $10^{16}$ &  1938  &  \textbf{224.9} &  589.2 &  422.2 &  167.0 &  0.38  \\
 74 &  12 &  $10^{12}$ &  9685  &  941.5 &  \textbf{861.7} &  4.6 &  857.1 &  1.09  \\
 61 &  34 &  $10^{27}$ &  10084 &  1661.3 &  \textbf{1579.5} &  139.5 &  1439.9 &  1.05  \\
 73 &  21 &  $10^{24}$ &  13449 &  1067.7 &  \textbf{1024.0} &  5.0 &  1018.9 &  1.04  \\    
        \end{tabular}
        }
    \end{center}
\end{table}

\begin{itemize}
\item The computation for BioModel 183 took almost one hour
to preprocess. This is due to the extraordinary 
number of polyhedra in some input bags.

\item 
Taking the time for preprocessing into account,
models with more than 800 polyhedra as solution---with the exception of BioModel~103---
gained a moderate speed-up (4\%--18\%) through preprocessing.

\item Preprocessing increased the overall run-time for 
all models with less than 400~polyhedra as solution, sometimes 
by a factor of 10 or more.
\end{itemize}

\subsection{Profiling of the Different Parts}

The relative time needed in different parts of the procedure 
to compute the whole solution
varies with the number of resulting polyhedra.  Table~\ref{tab:relative}
shows the relative time spent in three different parts of the procedure:
\begin{itemize}
\item \texttt{S}: Searching for another point outside the already known polyhedra,
\item \texttt{M}: Minimizing the found polyhedron,
\item \texttt{I}: Inserting that polyhedron into the list of already known ones and testing for inclusion.
\end{itemize}

\begin{table}[!ht]
    \caption{Relative run-time used for \texttt{S}earching another point, 
    \texttt{M}inimizing a polyhedron and \texttt{I}nclusion checking for models
    that took more than one second to compute, sorted by ascending number of polytopes.
    The second part of the models did not finish and the numbers signify
    only the times until the process was terminated. The largest portion of time
    is set in boldface.}
    \label{tab:relative}

    \begin{center}
        {\scriptsize
        \begin{tabular}{rrr|rrr}
        BM & PH & Time & \% in \texttt{S} & \% in \texttt{M} & \% in \texttt{I}  \\
        \hline\Tstrut
        183 &  1     &     77.7 &  \textbf{91.7} &  4.3 &  0.1  \\
638 &  13    &      3.2 &  \textbf{63.6} &  27.1 &  1.3  \\
200 &  20    &      2.2 &  \textbf{52.8} &  35.9 &  1.5  \\
599 &  24    &      1.6 &  28.0 &  \textbf{56.2} &  3.0  \\
489 &  42    &      3.0 &  31.6 &  \textbf{55.2} &  2.9  \\
147 &  54    &      4.4 &  34.2 &  \textbf{53.3} &  2.9  \\
396 &  54    &      4.2 &  19.1 &  \textbf{62.5} &  6.3  \\
666 &  64    &      2.2 &  16.9 &  \textbf{60.2} &  6.8  \\
230 &  68    &      5.1 &  33.5 &  \textbf{53.3} &  4.4  \\
365 &  70    &      4.3 &  31.3 &  \textbf{55.0} &  3.5  \\
328 &  86    &      1.7 &  16.9 &  \textbf{56.8} &  5.7  \\
105 &  130   &      9.7 &  43.7 &  \textbf{45.8} &  3.1  \\
 22 &  147   &      1.8 &  19.1 &  \textbf{45.4} &  5.8  \\
431 &  155   &      6.6 &  22.3 &  \textbf{60.2} &  6.0  \\
222 &  192   &      3.1 &  27.6 &  \textbf{40.6} &  7.2  \\
407 &  212   &     15.7 &  13.8 &  \textbf{67.2} &  11.9  \\
498 &  214   &      6.1 &  19.4 &  \textbf{53.2} &  12.1  \\
 32 &  244   &     12.8 &  11.6 &  \textbf{62.2} &  17.2  \\
102 &  322   &     12.2 &  26.9 &  \textbf{51.9} &  8.9  \\
477 &  467   &     33.8 &  9.0 &  \textbf{51.9} &  30.5  \\
 93 &  596   &     55.3 &  15.7 &  \textbf{64.3} &  13.3  \\
576 &  756   &     48.6 &  12.7 &  \textbf{60.1} &  19.7  \\
501 &  916   &    102.8 &  11.5 &  \textbf{68.2} &  15.5  \\
430 &  1676  &    128.2 &  10.2 &  \textbf{50.8} &  32.1  \\
103 &  1938  &    224.9 &  27.5 &  \textbf{54.5} &  12.6  \\
 74 &  9685  &    941.5 &  4.5 &  17.8 &  \textbf{71.6}  \\
 61 &  10084 &   1661.3 &  7.9 &  31.5 &  \textbf{56.4}  \\
 73 &  13449 &   1067.7 &  4.2 &  15.9 &  \textbf{73.4}  \\
        \hline\Tstrut
         14 &  >\,2663 &  >\,7200.0 &  11.9 &  \textbf{73.1} &  14.5  \\
151 &  >\,7734 &  >\,7200.0 &  1.9 &  22.5 &  \textbf{74.8}  \\
410 &  >\,10612 &  >\,7200.0 &  5.6 &  40.5 &  \textbf{52.6}  \\
560 &  >\,12577 &  >\,7200.0 &  1.9 &  29.3 &  \textbf{67.5}  \\
730 &  >\,15310 &  >\,7200.0 &  4.3 &  36.2 &  \textbf{57.5}  \\
        \end{tabular}
        }
    \end{center}
\end{table}
        
From the numbers it is obvious that the inclusion check time 
takes relatively more time as the number of polyhedra grows.

In contrast, SMT checking for every next point needs
relatively less time as the number of polyhedra grows
and the time needed for minimization is slowly getting less
prominent as well. 

It looks like the quadratic time inclusion check is the limiting
factor for models with many polyhedra.

\subsubsection*{Some Profiling of Polyhedral Inclusion Testing}

As was described in Sect.~\ref{procedure}, the test for inclusion
of a newly found polyhedron in the set of already found ones can
be sped up by testing containment of the included point $\vect{x}$
which was found by the SMT solver. 

Some cursory investigation shows that the number of full checks 
that still have to be done over the course of the whole computation
is about 0.26--0.68 times the number of polyhedra in the solution.

Yet, even though the constant is low, the run-time
of the inclusion test is quadratic and it becomes the most
dominant part of the computation for large result lists.

\section{Conclusion}
\label{conclusion}

We presented a novel method to compute tropical equilibria (resp.\ prevarieties)
from an input of sets of polyhedra. We sketched an algorithm for that 
purpose and discussed several possible improvements.
Furthermore, we ran extensive benchmarks with different SMT solvers 
to compute equilibria of 
tropicalizations of 46~different BioModels.

The conclusion is that the novel method is working and its computation
times compare favorably with a known algorithm using purely
polyhedral methods.  The run-times were always smaller for problems that would otherwise
take more than 58 seconds to compute, sometimes by a factor of 60 or more. 
The novel method has also the advantage to be an anytime algorithm, hence
it computes more parts of the solution given more time or 
computation power.  This is of importance if 
computation of the entirety of the solution is infeasible.

The CVC4 SMT solver was overall the fastest solver in this application, yet
Yices outperformed CVC4 for models that could be solved in less than 12~seconds.

Preprocessing the input yielded only moderate speed-ups and only on models
that had more than 800~polyhedra as solution.  Conversely, preprocessing
was always more costly for models with less than 400~polyhedra.

\subsection{Future Work}

There is obvious potential for a parallel implementation of the procedure.
We should expect an almost linear speed-up.

The rising percentage of time spent checking for
inclusion of already known polyhedra should be addressed.  
If one could assign one-dimensional properties (like dimension) to polyhedra, 
only parts of the list of already known 
polyhedra would have to be checked for inclusion.

The computation of non-maximal polyhedra should be avoided. Hence
a good and fast heuristic for choosing the $P_i'$ to construct a 
polyhedron of high dimension would lower the number of rounds 
needed to compute the solution.

Another avenue is to find better and faster preprocessing to minimize
the problem.

\subsection{Acknowledgments}

This paper is written in grateful memory of the author's advisor Andreas Weber, who 
unexpectedly passed away recently.

The author thanks Thomas Sturm, who kickstarted his interest in SMT.
Furthermore, he thanks Jörg Zimmermann and Ovidiu Radulescu for
discussions and valuable input, and the three anonymous referees
for their helpful comments.

This work has been supported by the bilateral project
ANR-17-CE40-0036 / DFG-391322026 SYMBIONT.

\bibliographystyle{splncs04}
\bibliography{smtcut}

\end{document}